\documentclass[letterpaper,11pt]{article}

\usepackage{amsmath}
\usepackage{amssymb}
\usepackage{graphicx}
\usepackage{color}
\usepackage{fullpage}
\usepackage{wrapfig}
\usepackage{cite}
\usepackage{cancel}

\newcommand {\e}{\epsilon}
\newcommand {\g}{\gamma}

\newcommand{\ds}{\cancel}

\numberwithin{equation}{section}
\definecolor{Red}{rgb}{1,0,0}
\definecolor{Blue}{rgb}{0,0,0.8}

\title{Dark Light, Dark Matter and the Misalignment Mechanism}
\author{Ann E. Nelson and Jakub Scholtz}

\begin{document}

\maketitle

\begin{abstract}
We explore the possibility that the  dark matter  is a   condensate of a very light  vector boson. Such a condensate could be produced during inflation, provided the vector  mass arises via the Stueckelberg mechanism.   We derive bounds on the kinetic mixing  of the dark matter boson with the photon,  and point out several  potential signatures of this model.
\end{abstract}

\section{Introduction}

In the past decades there has been mounting evidence  that approximately 20\% of the energy density in the universe is nonbaryonic,   pressureless, and very weakly interacting. Structure formation, baryon acoustic oscillations, galactic rotational curves, the Bullet cluster -- all of these theories or observations point towards nonbaryonic invisible stuff, with the same equation of state as nonrelativistic matter. Unfortunately, there is no compelling experimental evidence that dark matter has any nongravitational interaction with the standard model. At present there are two large classes of dark matter theories  which are motivated by compelling solutions to other problems beside dark matter: Axions and various variations on Weakly Interacting Massive Particles. Both are potentially detectable using specific techniques. It is however possible that neither is correct. Amongst some of the less canonical  candidates lie several variants of  light massive vector particles \cite{Bullimore, Goodsell,Abel}. Quite a few authors have already considered some of the consequences of the existence of such particles \cite{Redondo,Mirizzi,Jaeckel,Abel,Ahlers,Holdom}. However, the conclusion of these authors is that it is difficult to obtain sufficiently cold dark matter from a light vector particle.

We propose a variation that allows us to generate an extremely cold light vector component of the Universe, with a pressureless equation of state. Our inspiration comes from the fact that nature seems to make use of almost every renormalizable Lagrangian term: Abelian and Non-Abelian gauge theories, Yukawa couplings, $\phi^4$ theory, and yet even though the Stueckelberg mass \cite{Stueckelberg, Feldman} is renormalizable in 4 dimensions it does not appear in the standard model. If nature indeed does not make arbitrary choices among consistent theories then there should exist a spin one field with a Stueckelberg mass. Therefore, if this boson is sufficiently weakly coupled to the Standard Model,   it is worth considering whether it is a good candidate for dark matter. It is easy to populate the Universe with this particle: like the axion\cite{Fox}, during inflation, the expectation value of a light boson fluctuates. Immediately after inflation the value of the field in our horizon is a randomly selected (or perhaps anthropically selected \cite{Linde:1987bx,Turner:1990uz,Tegmark:2005dy}) initial condition. After inflation, when the Hubble constant is of order of the boson mass, the field begins to oscillate. This oscillating field may the thought of as a Universe-sized Bose-Einstein condensate, as described in section \ref{MM}.  Such a particle is allowed to kinetically mix with the photon via a renormalizable interaction. Therefore, at some level, it presumably does mix, although  no lower bound on the mixing parameter is required for the model to work. In section \ref{LIMS} we find the upper bounds on the kinetic mixing parameter such that the Early Universe  neither thermalizes nor evaporates this condensate. We also ensure that the vector boson lifetime is sufficiently long, and consider constraints on the coupling from possible apparent Lorentz violating effects.

\section{A Model of Light Vector Dark Matter}
Our massive vector will be represented by $\phi^\mu$ in a Lagrangian of the form:
\begin{equation}
 -\mathcal{L} = \frac{1}{4}\left(F^{\mu\nu}F_{\mu\nu}+\phi^{\mu\nu}\phi_{\mu\nu}+2\chi\phi^{\mu\nu}F_{\mu\nu}\right)+\frac{M^2}{2}\phi_\mu\phi^\mu+J_\mu A^\mu
\end{equation}
where $A^\mu$ and $F^{\mu\nu}$ represent the field strength of ordinary photon, $J^\mu$ is the ordinary charged current and $\phi^{\mu\nu} = \partial^\mu\phi^\nu-\partial^\nu\phi^\mu$. Applying a non-unitary transformation ($A\rightarrow A-\chi \phi $ and $ \phi \rightarrow \phi+\mathcal{O}(\chi^2)$) we can redefine our fields in terms of the mass eigenstates called \textit{massless} photon and \textit{heavy} photon:
\begin{equation}
 -\mathcal{L} = \frac{1}{4}\left(F^{\mu\nu}F_{\mu\nu}+\phi^{\mu\nu}\phi_{\mu\nu}\right)+\frac{M^2}{2}\phi_\mu\phi^\mu+J_\mu ( A^\mu-\chi\phi^\mu)
\end{equation}
By rotating ($\tilde{A} = A-\chi \phi $ and $\tilde{\phi} = \phi + \chi A $) we can reach the flavor eigenstates, called \textit{interacting} and \textit{sterile} photon. These two mix through their mass term:
\begin{equation}
 -\mathcal{L} = \frac{1}{4}\left(\tilde{F}^{\mu\nu}\tilde{F}_{\mu\nu}+\tilde{\phi}^{\mu\nu}\tilde{\phi}_{\mu\nu}\right)+\frac{M^2}{2}(\tilde{\phi}_\mu-\chi \tilde{A}_\mu)(\tilde{\phi}^\mu-\chi \tilde{A}^\mu)+J_\mu \tilde{A}^\mu
\end{equation}
Unless otherwise stated, we will use the mass eigenstate basis. In this basis, the heavy photon couples to the electromagnetic  current with the coupling constant scaled by $e \rightarrow \chi e$.  In the limit  $M=0$,  we could perform a rotation between the degenerate heavy and light eigenstates to a new set of states, one of which would be massless and completely decoupled. However, if we assume the heavy photon is the dark matter, with a finite energy density, produced via the misalignment mechanism, then the number density is inversely proportional to the mass and therefore it makes no sense to take this limit.

Note that the model has 2 free parameters: $M$ and $\chi$. Fundamental theory gives us little guidance for their values. The theory is technically natural for any values of $M$ and $\chi$, in the sense that for a cutoff of order the Planck scale, the renormalized values are of similar size to  the bare values. If we assume that the $U(1)$ of the standard model is grand-unified into a semi-simple or simple group, then $\chi$ can only be induced via loop corrections. In this case, if the mass of the particles in the loops $m_C$ is below the grand unification scale $\Lambda_{GUT}$,  the natural size of $\chi$ is of order $(g^2/(16\pi^2))^n$, where $n$ is the number of loops required to induce the kinetic mixing, and $g$ is the relevant combination of coupling constants in the loops. We will see that for $\phi $ to be   viable dark matter, $\chi$ has to be tiny, less than $10^{-7}$ over the entire mass range, so for $g\sim 1$, $n$ should be greater than or equal to about 3. If the particles in the loops are heavier than the grand unification scale, there is an additional suppression of at least  $(\Lambda_{GUT}/m_C)^2$.

\section{Misalignment Mechanism for Vector Dark Matter genesis during Inflation}
\label{MM}
The misalignment mechanism for producing a boson condensate has been considered in connection with the axion \cite{Preskill,Abbott,Dine,Turner,Fox} and various other light scalar fields such as moduli. Spatially varying modes of a bosonic  field will be smoothed by the expansion of the universe. However the zero-momentum component of the scalar field $A$ in the FRW background has the equation of motion:
\begin{equation}
\ddot{A}+3H(t)\dot{A}+m^2A=0
\end{equation}
which is reminiscent of harmonic oscillator, with a time dependent damping term $H(t)$. In the early Universe, $H(t)\gg m$, the scalar is effectively massless and its Compton wavelength does not fit into the horizon. The field is stuck: it does not go through a single oscillation and therefore we observe no particles.  The value of the field is assumed to take on some random nonzero value, because when the mass term is negligible there is no reason to prefer a field value of $\phi^\mu=0$. An episode of inflation will generally produce a spatially uniform field, but for $m\ll H$ in any causally connected patch of the universe the mean value of the field  takes on some random, non zero value. After inflation, the Hubble constant begins to decrease. As soon as the discriminant $9H^2-4m^2$ becomes negative, the field $A$ begins to oscillate and we can quantize the different modes and call them particles. Since, up to the small perturbations in the temperature, $H(\eta)$ is everywhere the same,  the transition happens everywhere in the Universe at the same time (in the rest frame of $A$). We are left  an energy density  which may be thought of as a coherent state  of a macroscopic number of particles. The particles are extremely cold and nonrelativistic, whatever their mass. An adiabatic perturbation spectrum arising from the fluctuations of the inflaton field \cite{Axenides:1983hj,Turner} will imprint adiabatic spatial variations on the density of the scalar particles, as is needed to fit the WMAP data. On large distance  scales compared with the particle Compton wavelength $1/m$ the dynamics  of gravitational structure formation is identical to that for any weakly interacting massive particle.  

Note that inflation will produce isocurvature perturbations arising from fluctuations of the scalar field $A$. Such perturbations are highly constrained, and will place an $m$ dependent upper bound on the inflation scale \cite{Turner,Lyth:1989pb,Turner:1990uz,Lyth:1992tx,Beltran:2006sq,Burns:1997ue,Fox,Hertzberg:2008wr} for this scenario.

We can show that the same scenario applies to a light massive vector in a FRW Universe. As shown in the appendix, the equation of motion for such a vector is:
\begin{equation}
\label{EOMFRW}
 -\partial_\nu \left(\phi^{\mu\nu}\sqrt{-g} \right)= - M^2 \phi^\mu \sqrt{-g}
\end{equation}
As  inflation blows up a small patch of space, we can assume the dark photon is uniformly distributed  and picks a particular polarization. This means that in the Cosmic frame $\partial_i\phi^\mu = 0$, and the time component of \eqref{EOMFRW} implies $\phi^0 = 0$ as long as $M \neq 0$. The spatial component of \eqref{EOMFRW} satisfies:
\begin{equation}
\label{EoM}
 \ddot{\phi^i}+3H\dot{\phi^i}+M^2\phi^i=0
\end{equation}
We see that each spatial component of the vector satisfies the same equation of motion as the scalar $A$ in the previous example and so has the same dynamics. After entering the lightly damped oscillation regime the vector behaves just like dust with $d(\rho a^{3})/dt=0$ where $\rho = \langle M^2\phi^2 \rangle$. Taking the upper bound of $\phi = m_{pl}$ when $M\sim H$ we can see that the mass of $\phi^{\mu}$ should satisfy $ M \geq  \Omega_{DM}^2 H_0 \hbar = 6.6\times10^{-35}\text{eV}$. This mass corresponds to a wavelength of about $10^{11} \text{pc}$. This lower bound on the mass is weaker than the one implied by the existence of compact galaxies \cite{vanDokkum} with $L \sim 1\  \text{kpc}$ and $M_{CG} \sim 2\times10^{11}M_{\odot}$. Requiring that the Compton wavelength of the dark matter is low enough to allow structure formation on the kpc scale gives a sharper bound on the lowest mass: 
\begin{equation}
1\:\mathrm{kpc} < \frac{\hbar}{\Delta p} = \frac{\hbar}{M v_{esc}}\;\; \Rightarrow \;\; M \geq 1.67\times10^{-24}\:\mathrm{eV}
\end{equation}

The amount of dark matter produced by this mechanism becomes simply a randomly chosen initial condition for the value of the field in our patch of the universe. In other regions of the universe, which are beyond our current horizon, the dark matter abundance is different. In ref. \cite{Tegmark:2005dy} it was shown, that for an axion or similar dark matter condensate produced during inflation, assuming other parameters do not vary, the regions of universe with dark matter abundance of order the abundance in our observed universe are the most highly correlated with physical features of our universe that seem favorable for existence of observers, allowing for an ``anthropic'' explanation of the dark matter density.

\section{Stueckelberg versus the Higgs mechanism}
The vector mass $M^2\phi^\mu\phi_\mu$ is not manifestly gauge invariant.  In the Standard Model of particle physics, all massive vector particles acquire their mass due to a Higgs mechanism. However, if the $\phi$ were to get its mass from the Higgs mechanism, the inflationary misalignment mechanism will not work to produce a condensate. Assuming the Higgs Lagrangian is:
\begin{equation}
\mathcal{L}  = [(\partial_\mu + ig \phi_\mu)\varphi]^2 + \lambda(\varphi^2-m^2/(2\lambda))^2
\end{equation}
the mass term for $\phi$ is $M^2\phi^2 = g^2m^2\phi^2/(2\lambda)$, however, the symmetry breaking happens around $T^2 \sim m^2/g^2$, which implies that the $\phi$ is massless above this temperature. Therefore, in order to make sure that there exists a time when $M \leq H(T)$ while $\phi$ is not massless, we need to satisfy:
\begin{equation}
 1 \leq H/M = \frac{T^2}{Mm_{pl}} = \frac{m^2}{Mg^2 m_{pl}} = \frac{2\lambda M}{g^4 m_{pl}} 
\end{equation}
Therefore we need:
\begin{equation}
\label{cond}
 \frac{M}{m_{pl}} \geq \frac{g^4}{2\lambda}
\end{equation}
We can look at the Z boson to illustrate this condition: the right hand side is of the order $2g^4v^2/m_h^2 \sim 10^{-3}$ even for a  heavy Higgs ($500$ GeV) and so a condensate of W and Z bosons could not have been created by a misalignment mechanism. However, one could imagine taking the limit in which $M^2 = g^2v^2 = g^2 m^2/2\lambda$ is fixed, but both $m_h \rightarrow \infty$ and $\lambda \rightarrow \infty$. In this case the right hand side of \eqref{cond} can be made arbitrarily small and the vector retains its mass for arbitrarily high temperature. The limit $m_h \rightarrow \infty$ can be handled in a better way: parametrize the Higgs in polar coordinates $\varphi = (v+h)e^{i \theta /v}$ and integrate out the heavy $h$. The effective Lagrangian of the light degrees of freedom takes the form:
\begin{equation}
 \mathcal{L} =-\frac{1}{4}F^2-\frac{1}{2}(MA^\mu+\partial^\mu\theta)^2
\end{equation}
which is identical to the Stueckelberg Lagrangian \cite{Stueckelberg, Feldman}, with $\theta$ filling the role of the Stueckelberg scalar field which fixes the correct number of degrees of freedom for a massive vector. This Lagrangian is still invariant under:
\begin{align}
 \Delta_\lambda A &= A + \partial\lambda\notag\\
 \Delta_\lambda \theta &= \theta-M\lambda
\end{align}
A redefinition $\phi^\mu = A^\mu+\partial^\mu\theta$ leads to $F^{\mu\nu} = \phi^{\mu\nu} = \partial^\mu \phi^\nu - \partial^\nu \phi^\mu$ and gives us a massive vector described by:
\begin{equation}
 \mathcal{L_S} =-\frac{1}{4}\phi^{\mu\nu}\phi_{\mu\nu}-\frac{M^2}{2}\phi^\mu\phi_\mu
\end{equation}
Naturally, this Lagrangian is still invariant under $\Delta_\lambda$, although it is not invariant under the naive gauge transformation $\phi^\mu \rightarrow \phi^\mu+\partial^\mu\lambda$. Unlike the nonabelian case, for a $U(1)$ gauge theory, the Higgs boson is not needed to unitarize the scattering of the longitudinal mode of a massive vector boson, and is   unnecessary for renormalizability.

\section{Bounds}
\label{LIMS}
\subsection{Early Universe - Compton Evaporation}

In order to be a successful dark matter candidate, the dark photon has to be a stable particle both in vacuum and in the dense, ionized early Universe. For light $\phi$ bosons, we need to ensure that the dark photon population does not get thermalized, otherwise it would become ultra relativistic and fail to be a good dark matter candidate. As with photons and plasmas, the main process for thermalization is the Compton-like scattering process: $\phi e^\pm \rightarrow \phi e^\pm$. However, this process will be suppressed by a factor of $\chi^2$ with respect to two other processes: $\phi e^\pm \leftrightarrow \gamma e^\pm$. We will call the right going process Compton evaporation and the left going Inverse Compton evaporation. Therefore in order to ensure there are enough dark photons left after interaction with plasma, we need to require that Compton evaporation rate $\Gamma$ is smaller than the expansion rate of the universe $H(T)$. Such condition will also imply that the thermalization rate from Compton-like scattering will be small and we will be left with enough cold dark matter to populate our Universe. In order to investigate this bound we need to know the  product of the velocity and cross-section $v\sigma(M,p)$ as a function of the dark photon mass $M$ and electron three momentum $p$, which can be re-expressed for $M\ll m_e$:
\begin{equation}
\label{APS}
 v\sigma(M,p)  = \frac{8\alpha^2\chi^2 \pi (3m^2+2p^2)}{9m^2(m^2+p^2)}+\mathcal{O}(M)
\end{equation}
The width of the dark photon in plasma is then given by the thermal average over the electron momentum density distribution for a given temperature of the Universe. We would like this width to be smaller than the characteristic expansion rate of the Universe at given temperature:
\begin{equation}
 H(T)>\Gamma(T) = \int dp^3 \sigma(M,p)v(p)n(p,T,\mu(T))
\end{equation}
Where we have used the exact $\sigma(M,p)$, not the approximate expression~(\ref{APS}), $n(p,T,\mu)$ is the Fermi-Dirac distribution with chemical potential $\mu$. We have chosen $\mu = 0$ for $T \gtrsim m_e$ and after $T$ drops below $m_e$ it was picked to be consistent with today's electron co-moving density. Given that the early Universe is growing less and less dense, the strongest bound on $\chi$ is in effect at the earliest time the dark photon is present, that is at the time when the misalignment mechanism kicks in at $M \sim H$. This guarantees that if the dark photon survives the first characteristic time period, then it will not evaporate anymore during the subsequent time. The condition $H\sim \Gamma$ does not guarantee this, but is a lower bound on such survival. We find it is unnecessary to consider other particles than the electron, since the contribution of all other charged particles with mass $m_i$ and charge $q_i$ will be suppressed by a factor $g_i(m_e/m_i)^2(q_i/q_e)^4$ which together with their suppressed thermal momentum distributions will make their contribution small. Likewise, it is unnecessary to consider other evaporation processes such as $\phi \gamma \rightarrow \gamma\gamma$ since they become important for dark photon mass of order $M = (m_{pl}m_e^2)^{1/3} \sim 10^{13}\;\mathrm{eV}$ - well above the range we consider in this paper. The bound imposed by Compton evaporation is plotted in Figure \ref{LIM} and labeled Early Universe. We would like to point out two features. When the Universe reaches temperatures of order $T \sim 0.1 m_e$, its free charge density significantly drops and the evaporation process becomes much less effective. This temperature marks the generation of dark photon with mass $M \sim T^2/m_{pl} \sim 10^{-18}$ eV, hence the sharp dip in the bound on $\chi$ in this region. On the other hand, since the cross-section starts dropping off when $\sqrt{s} \sim m_e^2/M$ and $\sqrt{s} \sim T$, we can estimate a change in the slope of the bound around $M \sim (m_e^4/m_{pl})^{1/3}$ which agrees with the observed dip at $M \sim 10^{-2}$ eV.

\subsection{Decays}
Apart from Compton evaporation, we can consider pure vacuum decay processes, which become significant once $M>m_e$ or $M>M_W$. Requiring that the dark photon is stable on cosmological timescales requires that $\sum \Gamma_i < H_0$:
\begin{center}
\begin{tabular}{l|c|p{5cm}}
Process & Width & Notes \\
\hline
&&\\
$\phi \rightarrow l^+l^-$ & $\Gamma_1 = \chi^2 \alpha \frac{M^2+2m_l^2}{2M^2}\sqrt{M^2-4m_l^2}$ & $M>2m_l$, Exact \\ 
&&\\
$\phi \rightarrow \nu \bar{\nu}$ & $\Gamma_2=\frac{\chi^2 \alpha^3}{16\pi}\left(\frac{M}{M_W}\right)^4\sqrt{M^2-4m_\nu^2}$ & Estimate of the loop process \\ 
&&\\
$\phi \rightarrow \gamma\gamma\gamma$ & $\Gamma_3=\frac{17 \chi^2 \alpha^4 M}{11664000\pi^3}\left(\frac{M}{m_e}\right)^8$ & See \cite{Redondo}, valid for $M<m_e$
\end{tabular}
\end{center}
The bounds imposed by these decays are plotted in Figure \ref{LIM}.

\subsection{Earth Detection}

Although $\phi_\mu$ does not satisfy  Maxwell's equations, its coupling to ordinary matter is the same as that of the photon $A_\mu$. Therefore, a nonzero $\phi_\mu$ will appear as a combination of electric and magnetic fields with strength suppressed by a factor of $\chi$. Such fields will be detectable in various precision experiments and it is our desire to quantify the expected phenomena as accurately as possible.

By our hypothesis, in the dark matter rest frame, $\phi_\mu = \delta_{\mu 3} A_3 \cos(Mt)$ and so it will mimic an electric field  $E_3 =\chi A_3 M\sin(Mt)$. Given that the local density of dark matter is $T_{00} = M^2 A_3^2/2 = 0.3 \:\mathrm{GeV}/\mathrm{cm}^3$, we can infer that the amplitude of the electric field will be 
\begin{equation}
E  = \sqrt{2\times0.3 \:\mathrm{GeV}/\mathrm{cm}^3/\epsilon_0} \approx 3300\chi\:\mathrm{V}/\mathrm{m} 
\end{equation}
However, there is no reason to believe that the dark matter rest frame is identical with the Earth frame and hence   we need to perform a Lorentz boost to the right frame. Given $\vec{\phi} = (\phi_x,\phi_y,\phi_z)\cos(Mt)$ and $\vec{v}$ - the velocity with respect to the dark matter rest frame, the B-fields in the Earth frame will be:
\begin{equation}
 \vec{B} = \vec{\nabla} \times L_{\vec{v}}(\vec{\phi}) = \gamma M\vec{v}\times\vec{\phi}\cos(\gamma Mt) 
\end{equation}
We should note, that at $v=0.001c$, $\gamma=1+\mathcal{O}\left(10^{-6}\right)$, and that $|M\phi|$ is the magnitude of electric field in the dark matter rest frame. Therefore the B-field is simply $\vec{B} = \vec{v}\times\vec{E}$ - precisely as expected.

\subsubsection{Attenuation}

If these fields are to be detected by Earth based experiments we need to check that the dark photon field is not screened by the atmosphere or by the many shields that experimental physicists put up in order to protect their experiments from stray electric and magnetic fields. In materials, bound electrons will only contribute to shielding if $M$ falls close to some energy gap of a kinematically allowed transition, however, even such transitions will be suppressed by factor of $\chi^2$. On the other hand, free electrons in metals will allow a continuum of transitions, that would lead to Compton evaporation effects. Therefore, we will treat the interaction of dark photons with materials as a wall penetration by weakly interacting particles, similarly to what we have done with the early Universe. The change in dark photon density will be proportional to:
\begin{equation}
 a  = \exp\left(-\int dx\: n(x)\sigma(M,v) \right)
\end{equation}
where $n(x)$ is the free electron density of the shielding material and $\sigma(M,Mv)$ is the Compton Evaporation cross-section and $v = 0.001c$ is the assumed velocity with respect to the local dark matter flow. Given that the respective average densities of free electrons in the ionosphere and copper are $n_{at}\sim3\times 10^{11}\; \textrm{m}^{-3}$ and $n_{\mathrm{Cu}}\sim10^{29}\; \mathrm{m}^{-3}$, it is clear that a whole column of $1000\;\mathrm{km}$ of atmosphere corresponds to a layer of metal about $10^{-12}\;\mathrm{m}$ thick, which is much less than any normal electric shielding of earthborne experiments hence we can disregard this contribution . Moreover, the early Universe bound on $\chi$ gives an attenuation length longer than 1 meter in a copper plate and once combined with the bounds from the next section the attenuation length is larger than $10^{10}\:\mathrm{m}$. 

\subsubsection{Atomic Physics}
The Stark effect associated with the background dark electric field would induce a shift in the ground state energy of a hydrogen-like atom of order $\Delta E_S = -m_e(3a_0^2eE_{d}/2\hbar)^2 \sim \chi^2\times 10^{-15}\:\mathrm{eV}$ which is 5 orders of magnitude smaller than the current limits \cite{Schwob}, even if $\chi = 1$. The Zeeman effect would produce a shift of $\Delta E_Z = 5\chi\times 10^{-13}\:\mathrm{eV}$. This is still too small to register. The advantage of the Zeeman effect is that it is first order in the fields, hence in $\chi$, which makes up for the fact that the magnetic field is suppressed by a factor of $v/c$. However, effects linear in fields go as $\cos(Mt)$, implying a zero time average, and so a search without prior knowledge of $M$ would be time consuming. However, in the region of small mass ($M\lesssim 10^{-22}\:\mathrm{eV} \sim 1$ year) the slow oscillations imply no need for averaging. In this regime the slowly changing background electric field would mimic a slow drift in $\alpha$. As an example we can take a system comprised of two clocks: one driven by two photon transition from $1s \rightarrow 2s$ in hydrogen and the other by the hyperfine transition in cesium. The major correction to the hydrogen clock rate comes as a Stark effect with a relative shift in the frequency that goes as 
\begin{equation}
\frac{\delta\omega}{\omega}=\frac{\Delta E_{1s} - \Delta E_{2s}}{E_{1s}-E_{2s}} = -\frac{\Delta E_{2s}}{E_{1s}-E_{2s}} = \sum_{n\geq 2} \frac{|\langle 2,0,0\left| e \mathcal{E}z\right| n,1,0\rangle|^2}{(E_{200}-E_{n10})(E_{200}-E_{100})}
\end{equation}
Notice that the $n=2$ term dominates the sum since the degeneracy of the 2s and 2p states is broken by the lamb shift with $\Delta 
E(2p-2s) \sim 10^{-6}\;\mathrm{eV}$, whereas the rest is on the order of $1\;\mathrm{eV}$. Therefore:
\begin{equation}
\frac{\delta\omega_{\mathrm{H}}}{\omega_{\mathrm{H}}}= \frac{0.55(\chi e \mathcal{E}a_0)^2 }{(\Delta E_{lamb})(E_{1s\rightarrow 2s})} \sim 4\chi^2\times10^{-10}
\end{equation}
Note that the Zeeman shift is identical for the 1s and 2s orbitals and so there is no contribution linear in $\chi$.

In cesium the Stark shift does not distinguish the states, but the Zeeman effect contributes by splitting the hyperfine triplet into three distinct levels, and induces a change in the clock frequency on the order:
\begin{equation}
\frac{\delta\omega_{\mathrm{Cs}}}{\omega_{\mathrm{Cs}}} = -\frac{\mu_e B}{\Delta E_{hyp}} \sim 1.5\chi\times 10^{-8}
\end{equation}
Clearly the cesium clock effect dominates for small $\chi$. Therefore, as $\mathcal{E}$ oscillates very slowly, the experiment sees a drift in $\delta\omega/\omega$ which could be (naively) interpreted as drift in $\alpha$ of the order:
\begin{equation}
 \frac{ \dot{\alpha}}{\alpha} = M\frac{\delta \omega}{2\omega} \sim 1.5\chi\times10^{-8}\left( \frac{M}{10^{-22}\;\mathrm{eV}}\right)\;\mathrm{year}^{-1}
\end{equation}
However, if the frequency of the oscillations is comparable to the time scales of an experiment, such as sampling rate and averaging times of individual data points, the sensitivity becomes more complicated. We pick \cite{Fischer} as a model example to illustrate our point. Fisher \textit{et al.} made measurements in June 1999 and February 2003, which, given the spacing between these two dates can be interpretted as two measurements separated by 44 months ($T=1320$ days), each averaged over roughly one month ($t_0=30$ days).  Therefore, the experiment should perceive a change in the value of the field equal to:
\begin{equation}
\delta\phi(\varphi_0) = \frac{\phi_0}{t_0}\left(\int_{t=0}^{t=T}dt \cos\left[M(t+T)/h+\varphi_0\right]-\int_{t=0}^{t=t_0}dt \cos\left[Mt/h+\varphi_0\right] \right),
\end{equation} 
where $\varphi_0$ is an unknown phase of the field. Performing the integral and factorizing gives us:
\begin{equation}
\delta\phi(\varphi_0) = \frac{4\phi_0}{Mt_0}\;\sin\left(\frac{MT}{2}\right)\sin\left(\frac{Mt_0}{2}\right)\sin\left(\varphi_0+\frac{M(t_0+T)}{2} \right)
\end{equation}
This means that for certain finetuned phases $\varphi_0 \sim -M(t_0+T)/2 $ the experiment could see nothing by simply being unlucky. However, we know that 95\% of time $\left|\sin \left(\varphi_0+M(t_0+T)/2 \right)\right| \geq \sin\left(0.05/4\times 2\pi \right) = 0.0785$ and so 95\% of time $\delta \phi$ is larger than:
\begin{equation}
|\delta\phi| \geq \left|\frac{4\phi_0}{Mt_0}\;\sin\left(\frac{MT}{2}\right)\sin\left(\frac{Mt_0}{2}\right)\sin\left(\frac{5\pi}{200}\right)\right|
\end{equation}
 
\begin{figure}[p]
\center{
\includegraphics{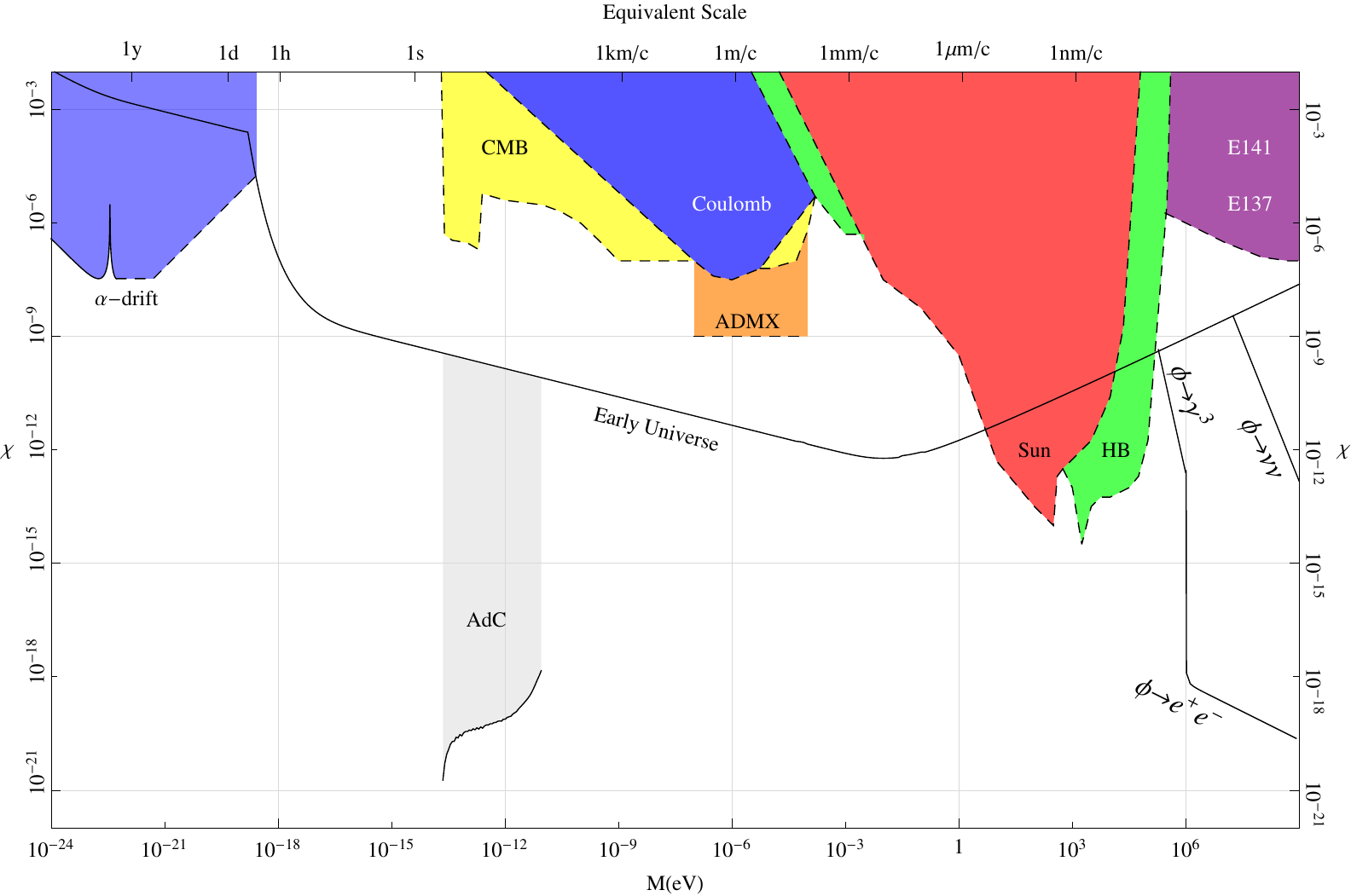}
\caption{Summary of Constraints: The early Universe behavior puts a dominant bound on $\chi$ in the higher mass range, for $M>2m_e$ the bounds are dominated by decays. The Shaded region called AdC marks the possible combinations of $(\chi,M)$ that could lead to adiabatic conversions. We have marked out the projection of the limits that can be achieved by ADMX \cite{ADMX}(Orange) - axion search experiment turned into a light shinning through the wall experiment. The bounds put by shaded regions with dotted lines come from a summary by \cite{Goodsell} and comprise the bounds by both theoretical and experimental considerations such as lifetime of the Sun (Red), Horizontal branch Star limits (Green), Coulomb law tests (Blue), CMB pollution by the dark photon (Yellow) and beam dump experiments E141 and E137 (Purple). In the low mass region the dominant bound comes from the drift of fine structure constant (blue, solid/dashed).}
\label{LIM}}
\end{figure}

We use this expression to put a 95\% confidence bound on $\chi$ and plot it as $\alpha$-drift in Figure~\ref{LIM}. Note that in the event that the sampling frequency of the expriment is a harmonic of the the oscillation frequency of the field, the experiment will also become insensitive to such a drift. This would show up as an oscillatory behavior in the bound on $\chi$ and we have replaced the region where these oscillations become too narrow to display with a dashed line in Figure~\ref{LIM}.

We would like to conclude the analysis of the fine structure constant drift bounds with two notes. First, as the cesium contribution dominates and the exact interaction of different atomic levels in cesium is beyond the scope of this paper, we would like to shelve this bound as tentative and in need of  focused treatment. Second, presence of dark matter in form of dark photon only mimics a drift in $\alpha$ and could be potentially resolved from an actual drift  if one were to measure different energy splittings which depend on different powers of $\alpha$.

\subsection{Adiabatic Conversion}

In the flavor basis, the dark and ordinary photon mix through the off-diagonal mass terms. In a thermal environment the mass matrix takes the form:
\begin{equation}
 \mathcal{M}^2 = \frac{1}{2}\left( \begin{array}{cc}
m_\gamma(x)^2+\mathcal{O}(\chi^2) & -\chi M^2 \\
 -\chi M^2 & M^2+\mathcal{O}(\chi^2) \end{array} \right)
\end{equation}
where $m^2_\gamma(x,t) = e^2n_e(x,t)/m_e$ is the plasma mass, which may depend on time or position. Should  the plasma mass be slowly varying then there could be  an adiabatic conversion between different states. Mirizzi \textit{et al.} explore this effect in the context of changing electron density in the Universe as it expands and distorts the CMB through an excess of converted dark photons \cite{Mirizzi}, and they offer a very useful comparison of this process to the neutrino MSW effect. We observe that this process could be much more severe in the environment of ionized gas that forms a significant portion of a typical cluster of galaxies.

\begin{figure}[p]
\center{
\includegraphics{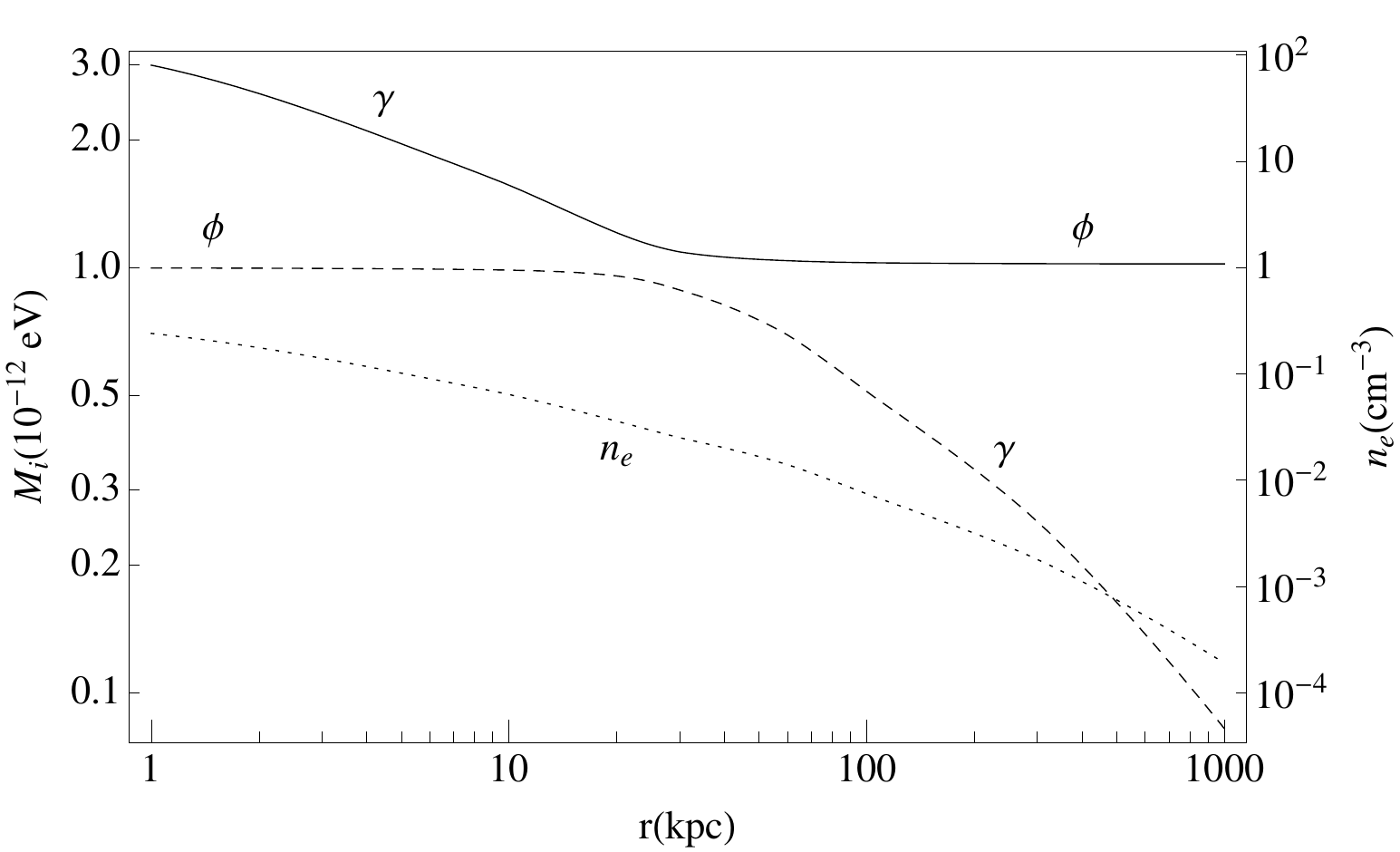}
\caption{Mass mixing in plasma: The solid and dashed curves show the eigenvalues of the mass matrix as a function of the radial position inside the cluster. The dotted line shows the density of the ionized gas in the cluster also as a function of the radial position. In order to make the level crossing visible we have adopted $M=10^{-12}\:\mathrm{eV}$ and $\chi=0.2$. }
\label{Mix}}
\end{figure}

Figure \ref{Mix} shows the energy of an eigenstate of the mass matrix as a function of radial distance of a particle from the center of the cluster. As an example we will follow a dark photon that is infalling into a cluster. If the dark photon infalls adiabatically, that is slowly enough, then it stays in the same eigenstate of the mass matrix which in fact contains more of the original photon state after it crosses the point where $m_\gamma \sim M$. Therefore, the dark photon is converted into an ordinary photon, which thermalizes very quickly (the cluster gas temperatures are in the range $10^6-10^7\;\mathrm{K}$, \cite{Markevitch}). Photons generate pressure and as a result the cluster loses its gravitational glue holding it together. Since we do observe clusters of ionized gas, it is imperative that the section of parameter space is excluded.

What does slow enough mean? In order to cross from one level to another we require that the characteristic time associated with the change in the system needs to be on the order of the gap between the energy levels. The rate of change of photon plasma mass close to the point where the energy gap is minimal is:
\begin{equation}
 t^{-1}|_{m_\gamma(x)=M} = \left.\frac{1}{m_\gamma(x)}\frac{dm_\gamma(x)}{dt}\right|_{m_\gamma(x)=M} = \left.\frac{v}{m_\gamma(x)}\frac{dm_\gamma(x)}{dx}\right|_{m_\gamma(x)=M} = \left.\frac{v}{2}\frac{n'(x)}{n(x)}\right|_{m_\gamma(x)=M}
\end{equation}
On the other hand the mass gap between the states is minimal when $M = m_\gamma$ and turns out to be:
\begin{equation}
 \Delta E|_{M=m_\gamma} = \chi M
\end{equation}
We take a free electron density curve from \cite{Croston}, replotted in \ref{Mix}, to determine the portion of $(\chi,M)$ parameter space in which this infall turns out to be adiabatic, taking the velocity of infall to be the escape velocity $v(r) = \sqrt{GM_c/r}$. We note that this mechanism will only work for a mass range of $10^{-13}$ eV - $10^{-11}$ eV, the lower limit coming from the density of voids and the upper from the highest densities inside clusters. We have plotted the resulting region in Figure~\ref{LIM} and marked it AdC.

\subsection{Breaking Lorentz Invariance}
The existence of dark matter necessarily causes apparent Lorentz violation because it defines a preferred frame - its own rest frame. The effects of this frame can be measured through its coupling to the standard model particles. However, even if those couplings were zero, in our case this corresponds to $\chi = 0$, there would be a gravitational interaction. Even in the dark matter rest frame there is  additional Lorentz violation due to the polarization of the dark photon.

Moreover, we can see the gravitational violation of Lorentz violation by looking at the stress-energy tensor: assuming the polarization points in the z-direction $A_i = \delta _{i3}A\cos(Mt)$, we get:
\begin{equation}
 T_{\mu\nu} = \frac{M^2A^2}{2}\left( \begin{array}{cccc}
-1 & & & \\
 & \cos(2Mt) & & \\
 & & \cos(2Mt)& \\
 & & & -\cos(2Mt)\end{array} \right)
\end{equation}
The time average of $T_{\mu\nu}$ corresponds to pressureless dust, just as we concluded from \ref{EoM}. Moreover, at late times the frequency of oscillations is shorter than the   expansion rate of the Universe.  The field begins to oscillate when $M\sim H$, and at this time the oscillations cannot be averaged over a period and the dark matter contribution to the Stress-Energy tensor is not rotationally invariant. However, at that early time  radiation dominates the energy density of the Universe and dark matter is a minor perturbation, therefore the lack of rotational symmetry of $T_{\mu\nu}$ does not produce any significant effect.

\section{Summary}
A nonrelativistic condensate of light vector particles could be produced during inflation and is a viable candidate for the dark matter component of the Universe. For ultralight vector particles, a small kinetic mixing term with the photon could allow this particle to be detectable. After considering the constraints on the mixing parameter from cosmology and astrophysics, we find that there are some   regions of parameter space which could give unusual laboratory signatures of dark matter, such as apparent time dependent shifts in electromagnetic properties of matter and dark matter conversion to visible photons in plasmas. It is peculiar that although Dark Photon picks a unique direction in the Universe, it lacks a mechanism to imprint this direction onto the Standard model contents of our Universe. This model offers a unique experimental signature - weak background electric and magnetic fields that cannot be screened. 

\section*{Acknowledgments}
JS would like to thank the following people for helpful discussions: Eric Adelberger, Guy Moore, Miguel Morales, Gray Rybka, Julianne Dalcanton and Tom Quinn. We acknowledge partial support from the Department of Energy under grant number DE-FG02-96ER40956.

\appendix
\section*{Appendix}
\section{Equations of motion in the Early Universe}

The kinetic term in the Lagrangian for a massive vector $\mathcal{L} = \phi^{\mu\nu}\phi_{\mu\nu}/4+M^2\phi^\mu\phi_{\mu}/2$, can be simplified to:
\begin{align}
 \frac{1}{4}\phi_{\mu\nu}\phi^{\mu\nu} & = \frac{1}{4}(\partial_\mu \phi_\nu-\partial_\nu \phi_\mu)(\partial^\mu \phi^\nu-\partial^\nu \phi^\mu)\notag\\
				 & = \frac{1}{2}(\partial_\mu \phi_\nu \partial^\mu \phi^\nu - \partial_\mu \phi_\nu \partial^\nu A^\mu)\notag\\
				 & = \frac{1}{2}(\partial^\alpha \phi^\beta)(\partial^\gamma A^\delta)(g_{\alpha\gamma}g_{\beta\delta}-g_{\alpha\delta}g_{\beta\gamma} )\notag\\
\frac{\delta \mathcal{L}}{\delta (\partial^\alpha \phi^\beta)} & = (\partial^\gamma \phi^\delta)(g_{\alpha\gamma}g_{\beta\delta}-g_{\alpha\delta}g_{\beta\gamma} ) = \phi_{\alpha\beta}\notag
\end{align}
Therefore, the equation of motion in curved space reads:
\begin{align}
 \partial^\alpha(\sqrt{-g} \phi_{\alpha\beta})=M^2 \phi_{\beta} \sqrt{-g}
\end{align}
Where in FRW metric this means:
\begin{align}
 -\partial_0 a^3(t)\phi_{0\beta}+a^3(t)\partial_i \phi_{i\beta} & =  a^3(t)M^2 \phi_{\beta} \notag\\
-3\dot{a}a^2 \phi_{0\beta}-a^3(t)\partial_0 \phi_{0\beta} + a^3(t)\partial_i \phi_{i\beta} & =  a^3(t)M^2 \phi_{\beta} \notag\\
-a^3(t)\left(\partial_0 \phi_{0\beta} + 3\dot{a}/a \phi_{0\beta} - a^2(t)\partial_i \phi_{i\beta} + M^2\phi_{\beta} \right) & =  0 
\end{align}
Keeping in mind that for our candidate $\partial_i \phi_\alpha = 0$ in the cosmic frame, $a(t) \neq 0$ after or during inflation and $\dot{a}/a=H$, the last line turns into:
\begin{align}
 \partial_0 \phi_{0\beta} + 3H \phi_{0\beta}  + M^2\phi_{\beta} &=0 \notag\\
 \partial_0 (\partial_0 \phi_\beta - \partial_\beta \phi_0) + 3H (\partial_0 \phi_\beta - \partial_\beta \phi_0) + M^2\phi_{\beta} &=0 
\end{align}
Therefore, the time component $\beta = 0$ gives us:
\begin{align}
  \partial_0 (\partial_0 \phi_0 - \partial_0 \phi_0) + 3H (\partial_0 \phi_0 - \partial_0 \phi_0) + M^2\phi_{0} &=0 \notag\\
M^2\phi_0 &= 0
\end{align}
On the other hand, the spatial component $\beta = i$ implies the equation for a Hubble-damped harmonic oscillator:
\begin{align}
 \partial_0 (\partial_0 \phi_i - \partial_i \phi_0) + 3H (\partial_0 \phi_i - \partial_i \phi_0) + M^2\phi_{i} &=0\notag\\
 \partial_0\partial_0 \phi_i + 3H \partial_0 \phi_i + M^2\phi_{i} &=0
\end{align}

\section{Compton Evaporation Matrix Elements}
For reference we have evaluated the matrix elements for the Compton Evaporation. The momenta were assigned as follows:
\begin{figure}[ht]
\center{
\includegraphics{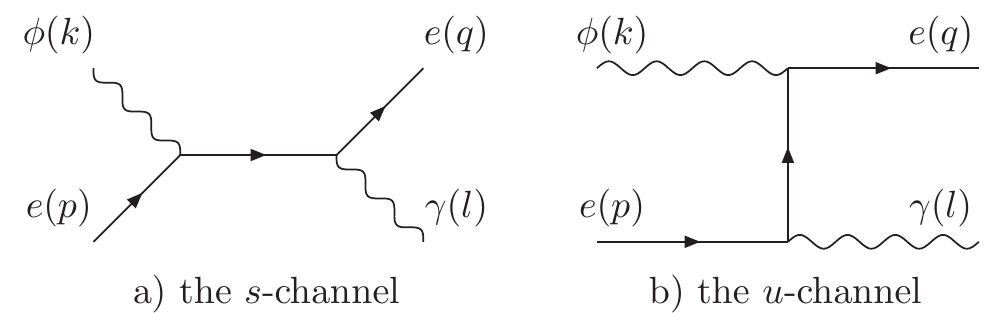}}
\label{DIAG}
\end{figure}
\\ 
With this convention the matrix element becomes:
\begin{equation}
iT = \chi e^2\e_\mu^*(l)\e_\beta(k) \bar{u}(q)\left( i \g^\mu \frac{-i(m-\ds{p}-\ds{k})}{(p+k)^2+m^2}i\g^\beta+ i \g^\beta \frac{-i(m-\ds{p}+\ds{l})}{(p-l)^2+m^2}i\g^\mu \right)u(p)\notag
\end{equation}
Which implies that:
\begin{align*}
\langle|T|^2\rangle=\frac{64\pi^2\chi^2\alpha^2}{3}&\left(2\frac{(m^2+M^2)(4m^2-t)-3m^2M^2}{(m^2-u)(m^2-s)}+\right.\\
&\left.+\frac{m^4+2m^2M^2+m^2(3s+u)-us}{(m^2-s)^2}+\frac{m^4+2m^2M^2+m^2(3u+s)-us}{(m^2-u)^2}\right)
\end{align*}
Which agrees, up to a number of polarizations factor, with regular Compton Scattering in the limit $M\rightarrow 0$, $\chi=1$.

\bibliographystyle{utphys}
\bibliography{DLDMBib}

\end{document}